\documentclass{jfm}
\usepackage{graphicx}
\usepackage{epstopdf, epsfig}

\usepackage{tabularx, booktabs, amsmath, subfig}
\usepackage[ruled,vlined]{algorithm2e}

\shorttitle{Deep Learning for Efficient Reconstruction of High-resolution Turbulence}
\shortauthor{Pranshu Pant, Amir Barati Farimani}

\title{Deep Learning for Efficient Reconstruction of High-resolution Turbulent DNS Data}

\author{Pranshu Pant\aff{1}
  \corresp{\email{barati@cmu.edu}},
  Amir Barati Farimani\aff{1}}

\affiliation{\aff{1}Department of Mechanical Engineering, Carnegie Mellon University,
Pittsburgh, PA 15213, USA
\aff{2}Department of Mechanical Engineering, Carnegie Mellon University,
Pittsburgh, PA 15213, USA}

\begin{document}

\maketitle

\begin{abstract}
Within the domain of Computational Fluid Dynamics, Direct Numerical Simulation (DNS) is used to obtain highly accurate numerical solutions for fluid flows. However, this approach for numerically solving the Navier-Stokes equations is extremely computationally expensive mostly due to the requirement of greatly refined grids. Large Eddy Simulation (LES) presents a more computationally efficient approach for solving fluid flows on lower-resolution (LR) grids but results in an overall reduction in solution fidelity. Through this paper, we introduce a novel deep learning framework SR-DNS Net, which aims to mitigate this inherent trade-off between solution fidelity and computational complexity by leveraging deep learning techniques used in image super-resolution. Using our model, we wish to learn the mapping from a coarser LR solution to a refined high-resolution (HR) DNS solution so as to eliminate the need for performing DNS on highly refined grids. Our model efficiently reconstructs the high-fidelity DNS data from the LES like low-resolution solutions while yielding good reconstruction metrics. Thus our implementation improves the solution accuracy of LR solutions while incurring only a marginal increase in computational cost required for deploying the trained deep learning model.
\end{abstract}

\section{Introduction}
Direct Numerical Simulation (DNS) is a highly accurate but expensive method for computationally solving the Navier-Stokes equations. Thus, even with the apparent fidelity benefits and the advent of modern computing, DNS has still not seen widespread adoption within the industry and the CFD community. On the other hand, Large Eddy Simulation (LES) which yields a spatially filtered approximation of the DNS results has much higher adoption within the industry and community in general. Solving on lower resolution grids like the ones used in LES, captures only the larger scale turbulent flows. It yields a coarse approximation of the solution and subsequently requires much less computation.  However, it also has some obvious drawbacks. LES solutions need augmentation via sub-grid scale (SGS) modelling in order to preserve unresolved small-scale physical processes. 
This results in a reduction in overall solution fidelity compared to DNS solutions. Herein lies the inherent trade-off between solution fidelity and computational complexity that CFD researchers and practitioners constantly grapple with. Through this paper, we suggest a novel machine learning (ML) based approach to mitigate this trade-off by ensuring high solution fidelity while keeping the computational overhead at a minimum.
Recently, machine learning especially deep learning has shown great promise in several interesting applications within the disciplines of computer vision and data driven modelling. Within fluid dynamics, deep learning has seen particular interest with applications to the field of turbulence modelling \citep{kutz2017deep}. Using machine learning, novel turbulence models have been developed for RANS simulations \citep{ling2016reynolds}. Additionally, machine learning has been used for developing closure models for CFD simulations \citep{beck2019deep, maulik2017neural, maulik2019subgrid, gamahara2017searching, bode2019deep}. ML architectures like convolutional auto-encoders (CAE) and Long Short-Term Memory (LSTMs) have shown great success in reduced order modelling (ROM) of fluid flows. Such architectures are replacing and augmenting conventional dimensionality reduction techniques such as Principal Orthogonal Decomposition (POD) and Dynamic Mode Decomposition (DMD) for developing reduced order models\citep{mohan2018deep, lee2020model, gonzalez2018deep}.
Through this paper, we aim to translate the novel advancements in machine learning to the domain of CFD. For this, we make use of deep neural networks to learn a mapping from the coarser LES like solutions to the refined DNS solutions. Such an ML network enables the reconstruction of high-frequency features that cannot be captured by the coarser low-resolution solutions.
Most importantly, by learning a complex and highly non-linear mapping between the low and high fidelity CFD simulations, massive savings in computational cost can be obtained. In fact, computational run times would decrease by orders of magnitude and would make the use of high fidelity CFD solutions more enticing. 
We framed the problem at hand to be analogous to up-scaling a low-resolution image to a high-resolution image (Super Resolution). So for our application, the result from a lower fidelity solver can be treated as the low-resolution (LR) input whereas the high-resolution (HR) output would approximate the DNS solution. 

Over the years, various ML architectures have continuously improved the Single Image Super Resolution (SISR) performance, particularly in terms of metrics such as Peak Signal to Noise Ratio (PSNR) and the Structural Similarity Metric (SSIM).
Some of the earliest works in SISR used a convolutional neural network (CNN) based approach such as SRCNN \citep{dong2015image}. Next, deeper CNN architectures were developed which incorporated residual blocks \citep{he2016deep} that significantly boosted the performance of conventional CNNs. Since then, several different implementations of residual blocks have shown improvement in certain aspects where previous models were shown to be lacking. DenseNets \citep{tong2017image} and Residual Dense Blocks(RDB)\citep{Zhang_2018_CVPR} are examples of this. More recently, Generative Adverserial Network (GAN) based methods like SRGAN \citep{ledig2017photo} and Enhanced SRGAN \citep{wang2018esrgan} have been able to be produce sharp and highly perceptually accurate images.
All the architectures described above are highly effective in image reconstruction but are often too heavily focused on attaining the best reconstruction metrics. Consequently, these architectures have large network sizes and are significantly lacking in terms of computational efficiency. Due to their large number of parameters, these networks consume a lot computational resources. Also, these large networks take really long to train and are notoriously hard to converge. These flaws renders these large networks unsuitable for use with CFD solvers where fast turnaround times are of the essence. So, in order to develop a framework for the efficient reconstruction of DNS data from LR data, an architecture is required which can yield decent reconstruction metrics while minimizing the network size at the same time.
Recently, ML models have also been used for upscaling of low-resolution CFD data. \citet{bode2019deep} and \citet{fukami2018super} are good examples of this. However, these implementations have shortcomings in certain key areas that we aim to address through our implementation. Reconstruction performed by \citet{bode2019deep} uses a GAN based model for upscaling which has several computational efficiency based downsides along with a propensity for non-convergence and difficulties during model training. The implementation by \citet{fukami2018super} utilises a shallow CNN architecture that doesn't make use of conventional deep learning techniques such as residual blocks. This severely limits the ML model's reconstruction capabilities and results in an overall reduction in reconstruction performance. Thus through this paper, we present SR-DNS Net, a deep learning framework that applies state-of-the-art super-resolution techniques to the domain of CFD. SR-DNS Net presents the possibility of upscaling low fidelity LR solutions to high fidelity DNS solutions in a computationally efficient manner. 
\section{Dataset}
\begin{figure}
  \centerline{\includegraphics[width=4in]{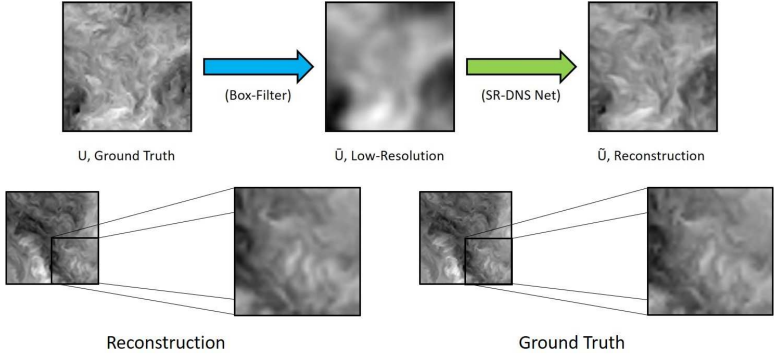}}
  \caption{Overview of data generation and reconstruction for SR-DNS Net}
\label{fig:dataset}
\end{figure}

\begin{figure}
  \centerline{\includegraphics[width=3.5in]{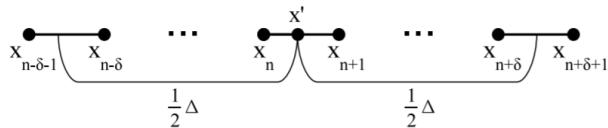}}
  \caption{Illustration for the box-filter Equation \ref{fig:dataset:box}}
\label{fig:dataset:box}
\end{figure}

For training our model and evaluating its performance we utilise the Isotropic Forced Turbulence data-set from the Johns Hopkins Turbulence Database (JHTDB) \citep{li2008public}.
This dataset contains direct numerical simulation (DNS) results for the Forced Isotropic Turbulence problem having a computational domain with $1,024^3$ nodes.
The Navier-Stokes equations for this problem are solved using the pseudo-spectral method and the Taylor-scale Reynolds Number for this simulation fluctuates around $\Rey \approx 433$.
In order to create low-resolution datasets, the DNS data is filtered using a box-filter (Equation \ref{eqn:dataset} and \textit{figure} \ref{fig:dataset:box}) of 3 different filter widths ($\Delta$). The filter width ($\Delta$) refers to the range of data points around the node under consideration that the box filter uses for filtering. The entire computational domain has 1024 nodes in each direction. For e.g. If the value of $\Delta$ = 21, the box filter uses 10 cells in the +x, -x, +y, -y, +z, -z directions of the grid point under consideration (\textit{figure} \ref{fig:dataset:box}). This results in the loss of high-frequency features and yields spatially filtered data that approximates coarse grid solutions. (\textit{figure} \ref{fig:dataset}). 3 different filter widths($\Delta=11, \Delta=21 \ \& \ \Delta=41$) are used to create a spectrum of low-resolution data. This helps in analysing and validating the model's reconstruction performance on a wide range of low-resolution data. Depending on the degree of refinement or ($\Delta$), the three datasets are referred to as Low-Resolution Fine (LR-Fine), Low-Resolution Medium (LR-Medium) and Low-Resolution Coarse (LR-Coarse).
\begin{gather}
  \overline{f(x')} = \frac{1}{\Delta^3}.\sum_{i=n-\delta}^{n+\delta} \sum_{j=n-\delta}^{n+\delta} \sum_{k=n-\delta}^{n+\delta} f(x_i,y_j,z_k)
\\
    where \ n = int(\frac{x'}{\Delta x} + \frac{1}{2}) = int(\frac{y'}{\Delta y} + \frac{1}{2}) = int(\frac{z'}{\Delta z} + \frac{1}{2}) \ and \ \delta = int(\frac{\Delta}{2\Delta x}) \notag
\label{eqn:dataset}
\end{gather}
The dataset has a Kolmogorov time scale of 0.0424s, time averaged over t=0 and 10.056 seconds. 3D velocity data is sampled from the entire computational domain of 1024 x 1024 x 1024 nodes and has periodic boundary conditions. This sampling is performed over a duration of 10.056 seconds with the data being queried at intervals of 0.1 seconds to allow for substantial differences in flow features between consecutive snapshots. The sampled data is then divided into separate channels of x, y and z velocities within 2 dimensional regions of 128 x 128 pixels. This segmentation is necessary for limiting the learnable parameters of the machine learning model to a manageable size given our computational constraints.
Each of the 3 final datasets consist of 50,000, 128 x 128 x 3 DNS and LR snapshots from the JHUTDB Forced Isotropic Turbulence Dataset. Finally, dataset augmentation is performed by randomly flipping the corresponding snapshots about the X and Y axes. This improves model training by increasing the number of training samples without storing additional CFD data in memory.
\section{Methodology}

To perform the reconstruction of the HR turbulent data we utilise a machine learning (ML) model which in its crux is just a function approximator($f$) with learnable parameters($\theta$). Given, the LR CFD data as input the machine learning model tries to generate a non-linear function mapping which can most closely approximate the HR CFD data as its output. The learning aspect of the model occurs via iteratively updating the function parameters. This update is based on the back propagation of the error between the ground truth ($y$) and the function prediction ($f(w,\theta)$). Thus, the problem reduces down to finding the function parameters $ \theta$ that can minimize the error between the ground truth and the prediction (Equation \ref{eqn:defn}). This paradigm of learning network parameters is often referred as supervised learning as the iterative learning is guided by comparison with the ground truth.
\begin{gather}
    \theta = argmin_\theta \ \epsilon (y, f(w,\theta))
    \\
    OR \notag
    \\
    \theta = argmin_\theta \ ||y-f(w,\theta)||_1 \notag
    \label{eqn:defn}
    \\
    Here, \ the \ error (\epsilon) \ is \ evaluated \ using \ the \ L1 \ loss \ metric \notag
\end{gather}
Given the similarity of our task with single image super resolution (SISR) problems, the ML models that performed well in the domain were examined. Moreover, an additional constraint in terms of computational cost was applied to narrow down the ML architecture subspace. Based on these constraints, Generative Adverserial Network(GAN) based SISR models were deemed infeasible. GAN based super-resolution models have large model sizes due to the presence of two independently trainable adversarial networks. These adversarial networks are referred to as generator and discriminator within the ML literature. Additionally, GANs and their propensity for training instabilities due to convergence oscillations, mode collapse and vanishing gradients has been well documented \citep{salimans2016improved}.
Given the above stated constraints and considerations, implementing deep CNNs with residual blocks (ResNets) emerged as the obvious choice. But again, the large number of trainable parameters associated with residual blocks severely bogged down the network's performance. Thus, we had to identify a CNN architecture with the capabilities of ResNets minus the massive computational overhead. Here, Mobilenets came to the fore by maintaining the network complexity while using only a fraction of the computational resources of ResNets \citep{howard2017mobilenets, sandler2018mobilenetv2}. This is because, MobileNets were developed to be used to be used for performing real-time image segmentation on embedded systems and mobile devices that are compute limited. To achieve this, MobileNets make use of the more efficient Inverted Residuals layers and Linear Bottlenecks instead of the regular Residual Blocks \citep{43022, chollet2017xception} and we did the same replacement for our framework.

\begin{figure}
  \centerline{\includegraphics[width=6in]{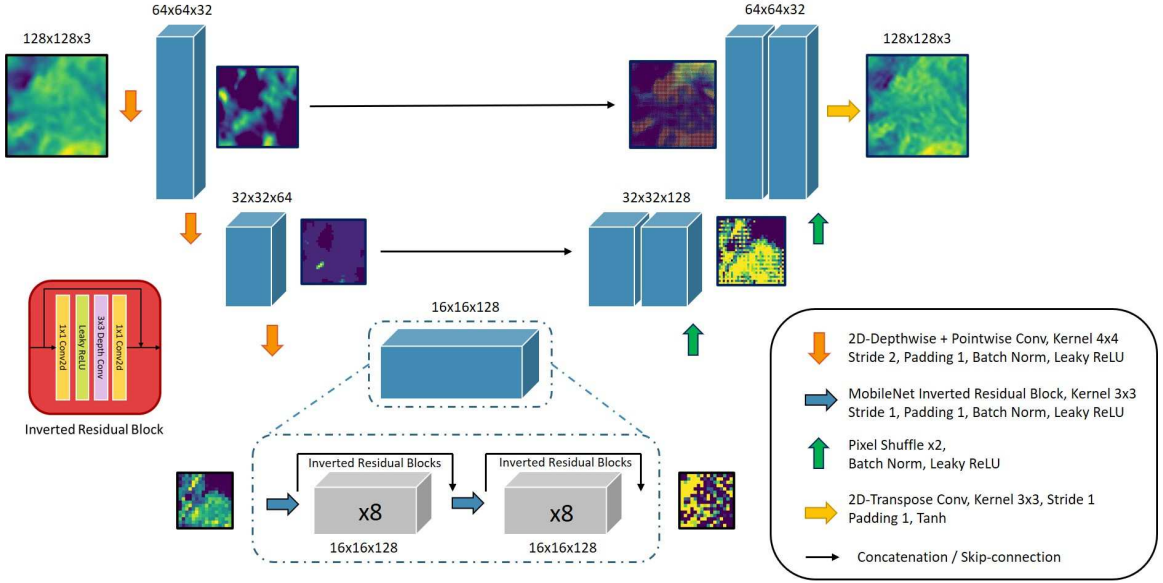}}
  \caption{Deep Learning network architecture for SR-DNS Net}
\label{fig:ka}
\end{figure}

Now, given the large input data sizes that we are dealing with (128 x 128 x3), we would not have been able to run the original MobileNet architecture in an efficient manner. So we explored the possibility of using an encoder-decoder architecture and finally decided to draw inspiration from U-Nets \citep{10.1007/978-3-319-24574-4_28}. U-Nets are a class of CNNs which have been used to achieve state of the art results in image segmentation tasks. These networks are referred to as U-Nets due to their distinct encoder-decoder architecture that takes on shape of a "U" (\textit{figure \ref{fig:ka}}). Using the U-Net's encoder layers we first downsample the input into a manageable dimension of 16 x 16. Only then we implement the 16 units of the MobileNet inverted residual layers. 

Nonetheless, in order to achieve a high-quality reconstruction a very deep neural network was still required. So, a network with 16 inverted residual blocks was implemented. However, such deep networks tend to encounter the problem of vanishing gradients which renders the ML models untrainable. To tackle this, we introduced a global skip connections within the architecture (represented by black arrows in \textit{figure} \ref{fig:ka}) and split the inverted residual blocks into 2 separate blocks of 8 with a skip connection in between(\textit{figure} \ref{fig:ka}). This ensured that the error was able to backpropagate through to the initial layers without diminishing significantly.

Finally, using the decoder of the U-Net we upsample the output of the inverted residual blocks to the required size. An important feature of U-Nets that greatly boost their performance are the skip-connections between corresponding layers of the encoder and decoder  (\textit{figure \ref{fig:ka}}). This further helps alleviate the vanishing gradient problem referred to earlier and also helps in improving the reconstruction performance.
Another important consideration in deciding the network architecture was deciding between 2D and 3D convolutions. While working with the forced isotropic DNS data, the convolution operation can be performed in the form of 3D convolutions across the data or by performing 2D convolutions on 2D slices(snapshots) of the velocity data. Performing 2D convolutions instead of 3D convolutions brings about a significant reduction in the number of trainable parameters, thereby reducing the computational cost. However, the downside of using 2D convolutions instead of 3D convolutions is the loss of spatial correlation in the direction that has been collapsed. In the end given our compute constraints, we decided to utilise 2D convolutions for our application.

Pixel-Shuffle was another state-of-the-art ML technique that we utilised for optimising the upscaling operation in our architecture. A Pixel-Shuffle layer \citep{shi2016real} provides a significant performance boost compared to a conventional deconvolution upsampling. 
Pixel-Shuffle is highly efficient in up-sampling images and can decrease the computational complexity of the network logarithmically in the image dimension compared to the conventional convolutional methods. Pixel-Shuffle achieves this boost in performance by learning to rearrange the superfluous pixels from the channel dimension into the height and width dimensions instead of performing expensive deconvolution operations. This upscaling by rearrangement compared to the expensive convolutions reduces the learnable parameters substantially. Thus, our final architecture is a MobileNet and U-Net hybrid which utilizes Pixel-Shuffle upsampling.

Additionally, we experimented with different loss metrics for evaluating the error. We found that the MAE(L1) loss yielded significantly better results compared to MSE(L2) for our use case. Moreover, the model trained on the L1 loss was found to be less susceptible to getting stuck in local minimas, something which was quite prevalent when using L2 as the loss metric. We believe that this difference in performance can be attributed the formulation of the loss metrics themselves. The slope of the L2 error is larger away from the origin but it quickly diminishes to zero near the origin. Whereas, L1 has a constant slope throughout its domain, even in the regions adjacent to the origin. Thus, using an L1 loss we can back-propagate error even when the error value approaches zero. Thus, we found that the L1 loss was more adept at yielding improvements even when the pixel-wise error was approaching zero.

\begin{figure}
  \centerline{\includegraphics[width=4in]{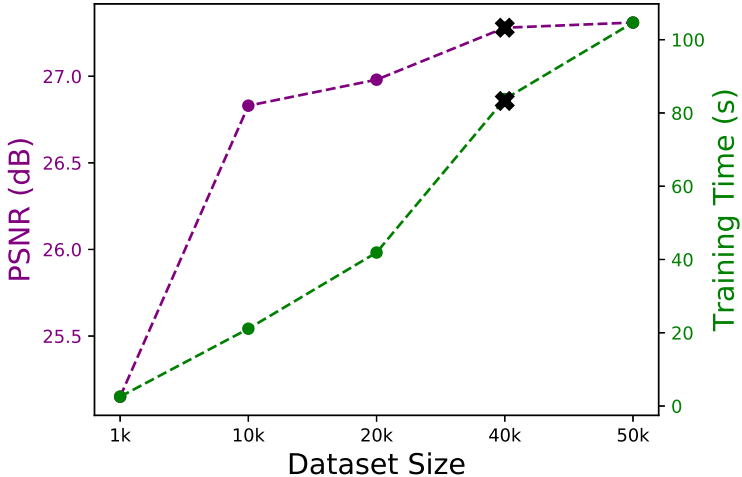}}
  \caption{Comparison of reconstruction performance v/s training time per epoch on the LR-Medium dataset}
\label{fig:dvt}
\end{figure}

Next, we performed a study to determine the effect of reducing the dataset size in the overall performance of the network. We ran the model using different sizes of the input training data ranging from 1k to 50k. We found that sampling 40k images from the 50k image dataset gave the most bang for its buck in terms of reconstruction performance and training times (\textit{figure \ref{fig:dvt}}).

Our final implementation uses the following dataset split. 40k snapshots for the train set, 5k snapshots for the validation split and 5k snapshots for the test set.
For training the model we use the Adam optimizer \citep{kingma2014adam} with an initial step size of 0.01 with a learning rate scheduler which reduces the step size by a factor of 10 whenever the model performance plateaus.
Additionally, during training we implemented early stopping to identify the most suitable model for reconstruction by reducing over-fitting. Finally, to train and deploy the deep learning model we used a system with an Intel Core I9-9900K processor with 16GB of RAM and an NVIDIA GeForce RTX 2080Ti GPU.

\section{Results}
To determine the efficacy of our ML model, we evaluated its performance on three datasets with varying levels of coarseness.
Comparison based on metrics such as Peak Signal to Noise Ratio (PSNR) and Structural Similarity Index Metric (SSIM) was carried out.
PSNR is the ratio of the peak signal power to the noise in the image. It is used to compare the quality of reconstruction in terms of pixel-wise accuracy. Since it is inversely proportional to MSE, higher PSNR values correspond to better reconstruction. SSIM is used to evaluate the perceptual or perceived similarity between two images $(\mathit{I}$ and $\mathit{\hat{I}})$ on a scale from 0 to 1 with 1 representing perfect perceptual similarity. 
The formula for these metrics can be seen in Equations \ref{eqn:metric-1} and \ref{eqn:metric-2}. The constant \textit{R} in equation \ref{eqn:metric-1} is the value of the highest pixel value in the image which is 1.0. $\sigma$ and $\mu$ represent the mean and standard deviation of the images under comparison in Equation \ref{eqn:metric-1}.
$\mathit{C_1}$ and $\mathit{C_2}$ are constants specifically chosen to prevent division by 0.

\begin{equation}
    \textit{PSNR}=10\log_{10}\left(\frac{R}{\textit{MSE}}\right)^2\\
    \label{eqn:metric-1}
\end{equation}

\begin{equation}
    \begin{split}
     \textit{SSIM}(I,\hat{I})&=\frac{(2\mu_I\mu_{\hat{I}} + C_1)(2\sigma_{I} \sigma_{\hat{I}} + C_2)}{(\mu_I^2 + \mu_{\hat{I}}^2 +C_1)(\sigma_I^2 + \sigma_{\hat{I}}^2 + C_2)}.
    \end{split}
    \label{eqn:metric-2}
\end{equation}

\begin{table*}
  \begin{center}
\def~{\hphantom{0}}
 \begin{tabular}{p{0.15\linewidth}p{0.125\linewidth}p{0.15\linewidth}p{0.15\linewidth}p{0.1\linewidth}p{0.125\linewidth}}
 \\
 \toprule
 \textbf{Dataset} & \textbf{Metric}  & \textbf{LR Input} & \textbf{Recon.} & \textbf{DNS} & \textbf{\%Improv.} \\ [0.5ex] 
 \hline
 LR Fine & PSNR  & 28.124 dB & 30.227 dB & $\infty$ & 7.477 \% \\ 
 $(\Delta = 11)$ & SSIM  & 0.912 & 0.926 & 1 & 1.502 \% \\
  & MSE  & 0.039 & 0.030 & 0 & 30.000  \% \\
 \hline
 LR Medium & PSNR  & 24.366 dB & 27.283 dB & $\infty$ & 11.971 \% \\ 
 $(\Delta = 21)$ & SSIM  & 0.828 & 0.850 & 1 & 2.680 \% \\
  & MSE  & 0.060 & 0.043 & 0 & 39.535 \% \\
 \hline
 LR Coarse & PSNR  & 21.364 dB & 25.030 dB & $\infty$ & 17.159 \% \\ 
 $(\Delta = 41)$ & SSIM  & 0.737 & 0.772 & 1 & 4.820 \% \\ 
  & MSE  & 0.085 & 0.056 & 0 & 51.786 \% \\
 \bottomrule
\end{tabular}
  \caption{Comparison of metrics between the LR Input, Reconstruction and Ground Truth(DNS) data on the test set}
  \label{tab:kd}
  \end{center}
\end{table*}

Using only reconstruction loss for backpropagating the error, our network significantly improves the PSNR and SSIM values of the LR data in all three datasets (\textit{table 1}, \textit{figure 3}). This super-resolution enhancement is also apparent on visual inspection of the reconstruction results as shown in \textit{figure 4}.
\begin{figure}
     \centering
     \subfloat[][PSNR]{\includegraphics[width=0.5\linewidth]{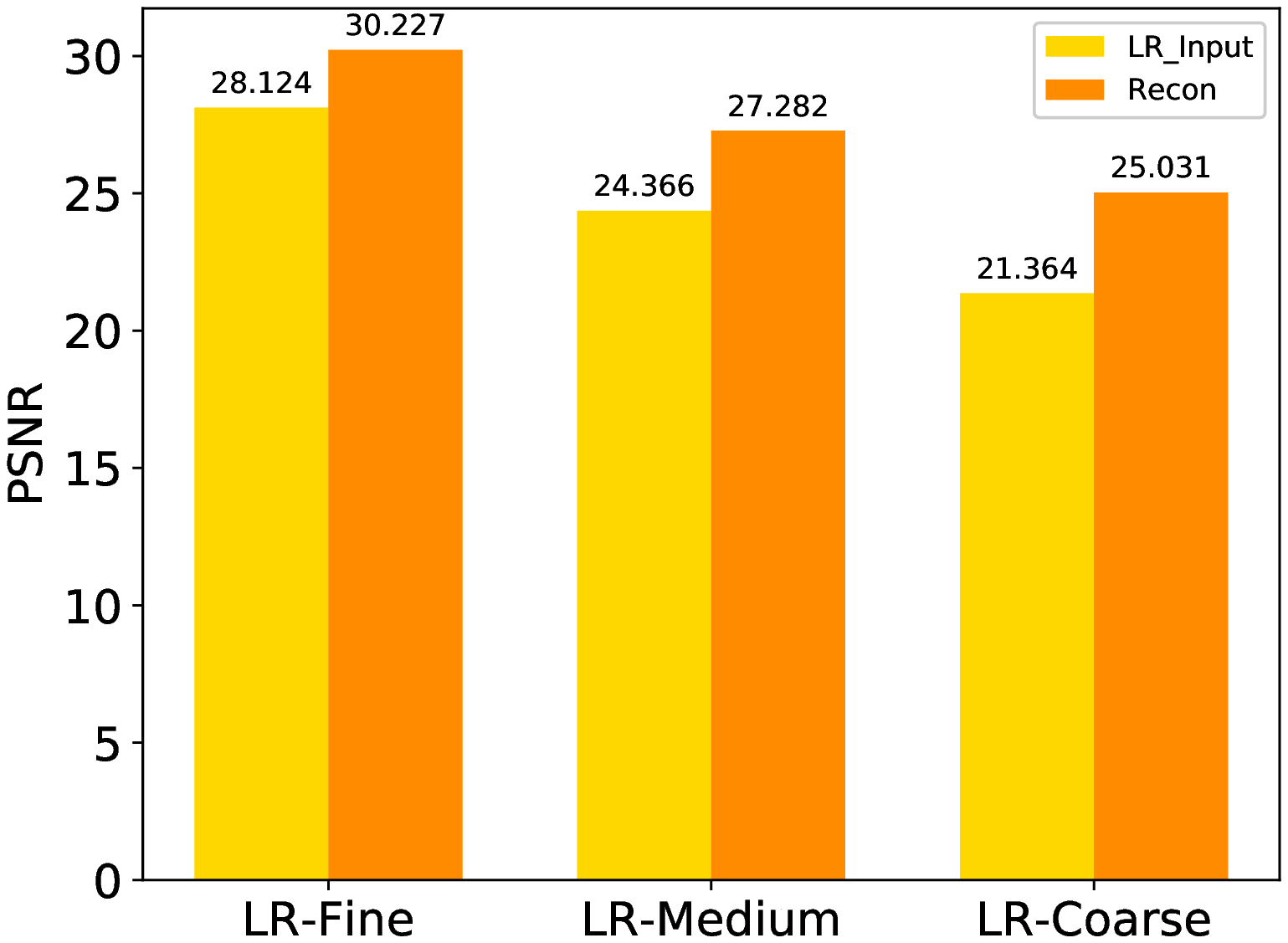}}\hfill
     \subfloat[][SSIM]{\includegraphics[width=0.5\linewidth]{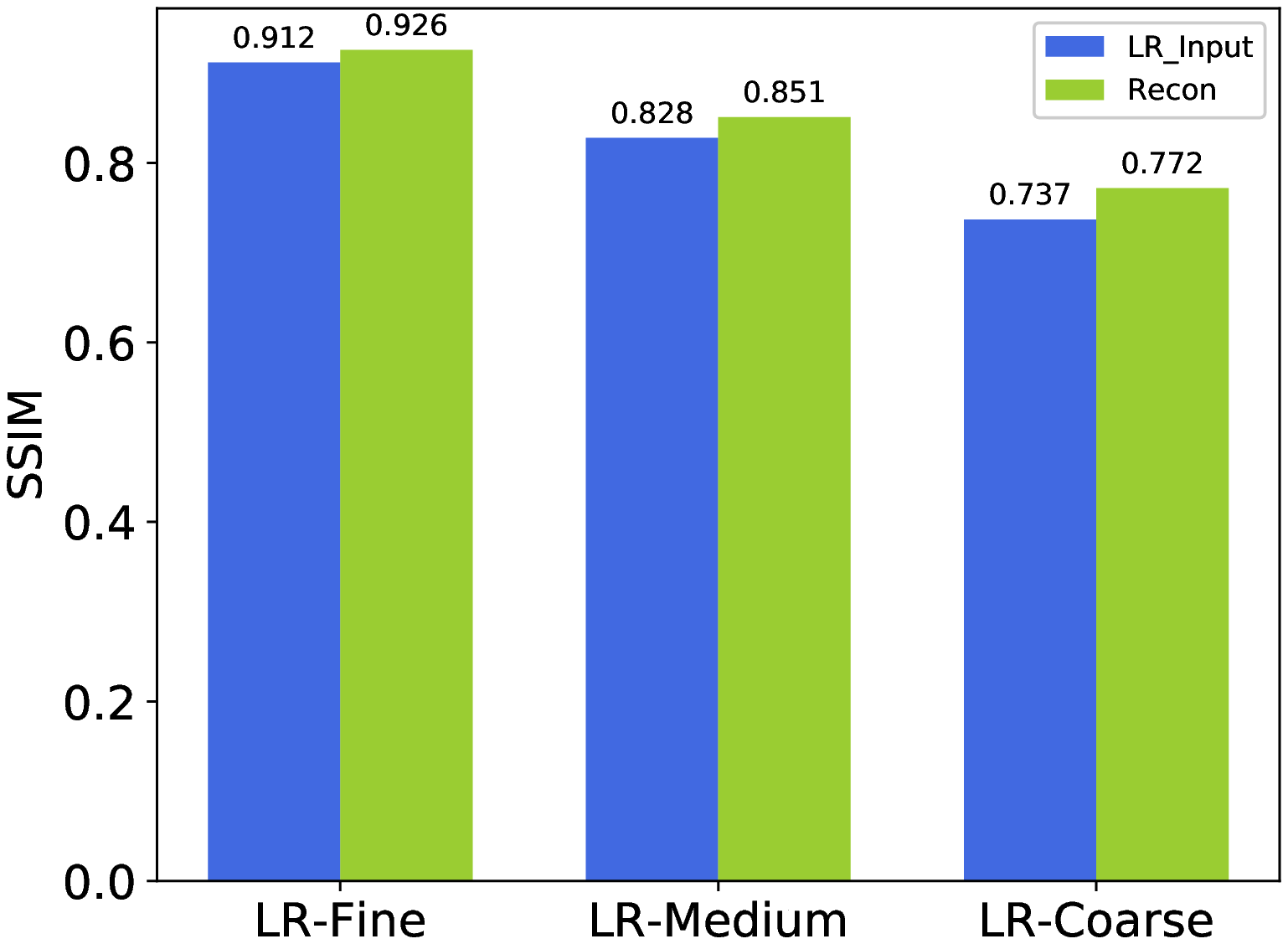}}
      \caption{PSNR(a) and SSIM(b) metrics for SR reconstruction on different LR datasets}
      \label{simulationfigure}
\end{figure}

\begin{figure}
     \centering
     \subfloat[][Normal Training]{\includegraphics[width=0.5\linewidth]{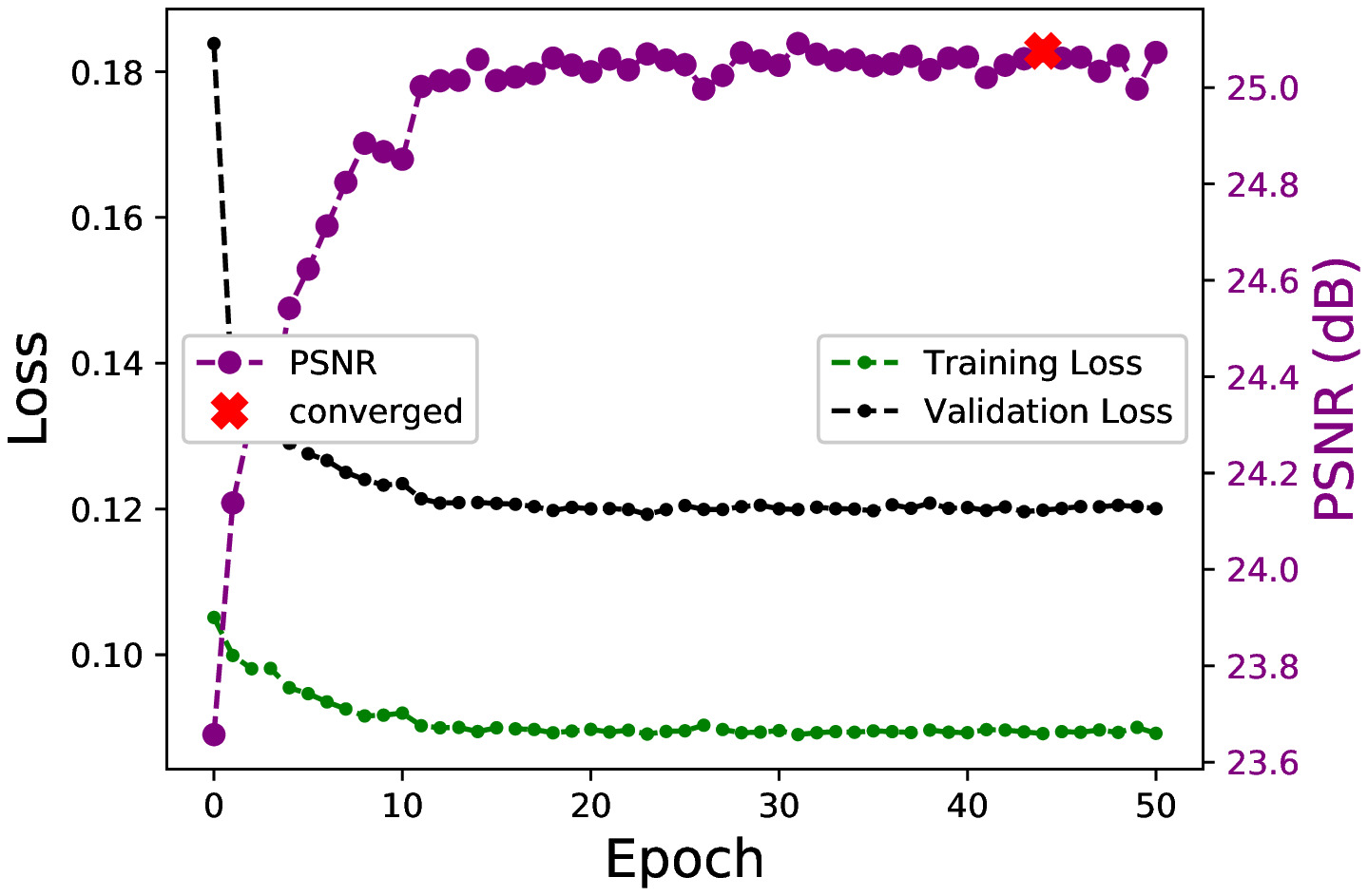}}\hfill
     \subfloat[][Transfer Learning]{\includegraphics[width=0.5\linewidth]{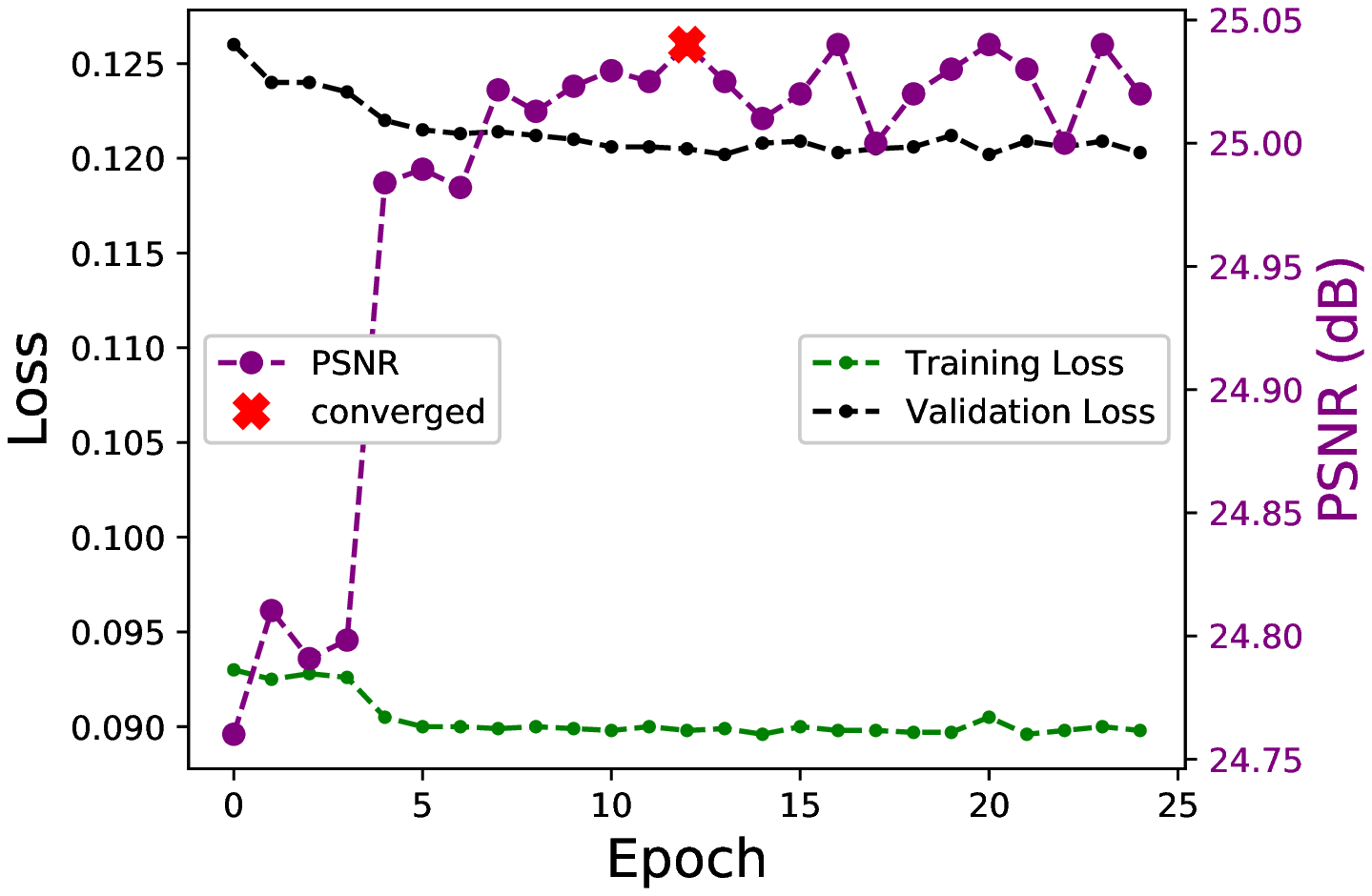}}
  \caption{Plot for training loss v/s epochs. a) represents the loss curves for normal training on the LR-Medium dataset. b) represents the loss curves with transfer learning on the LR-Coarse dataset with weights trained on LR-Medium dataset.}
  \label{loss_plot}
\end{figure}

\begin{figure}
\centering
     \subfloat[][LR-Fine Dataset]{\includegraphics[width=0.95\linewidth]{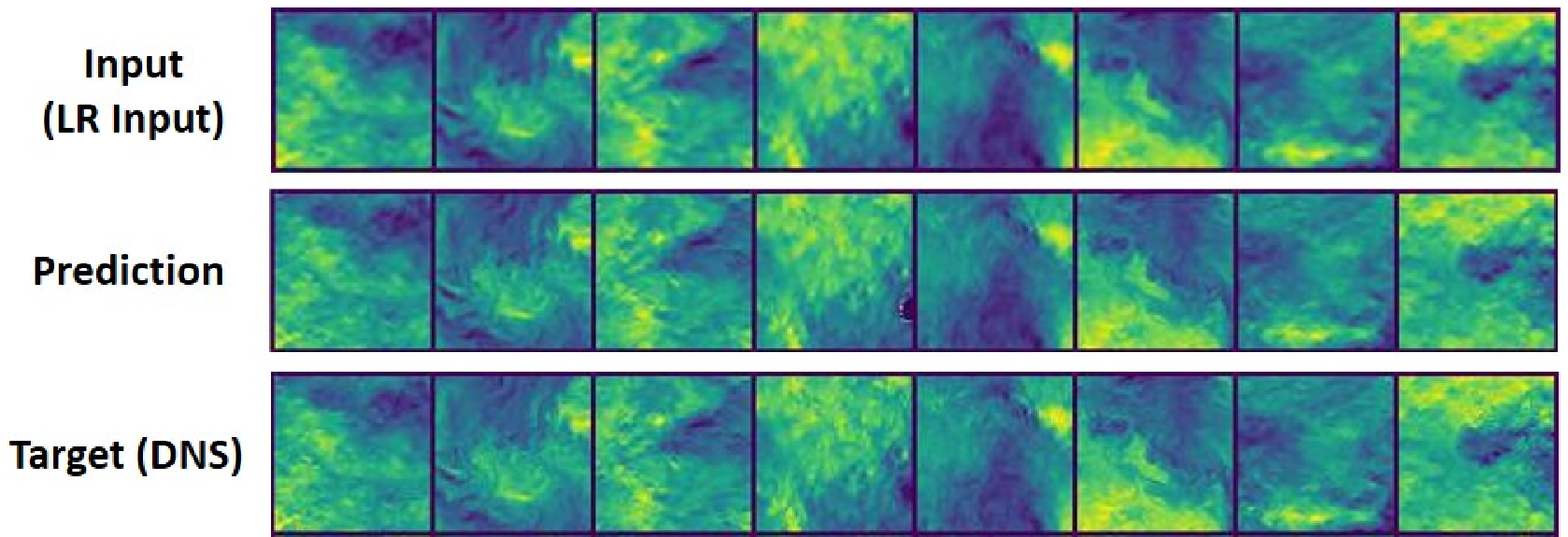}} \hfill
     \subfloat[][LR-Medium Dataset]{\includegraphics[width=0.95\linewidth]{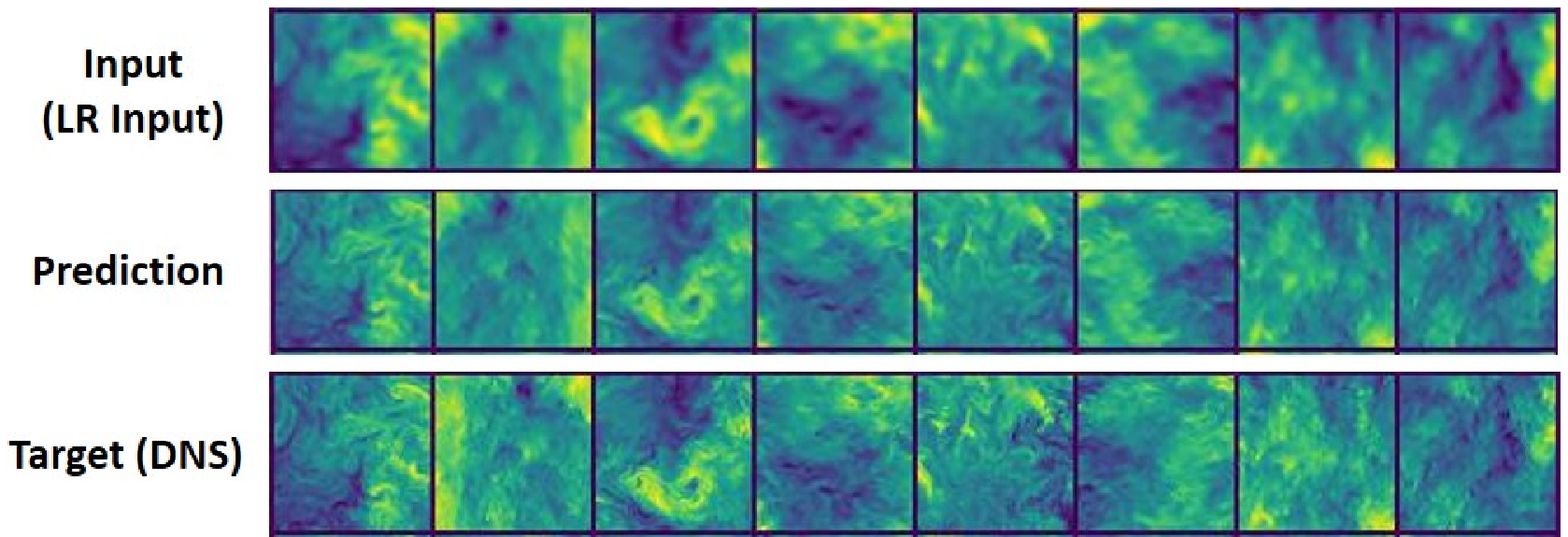}}\hfill
     \subfloat[][LR-Coarse Dataset]{\includegraphics[width=0.95\linewidth]{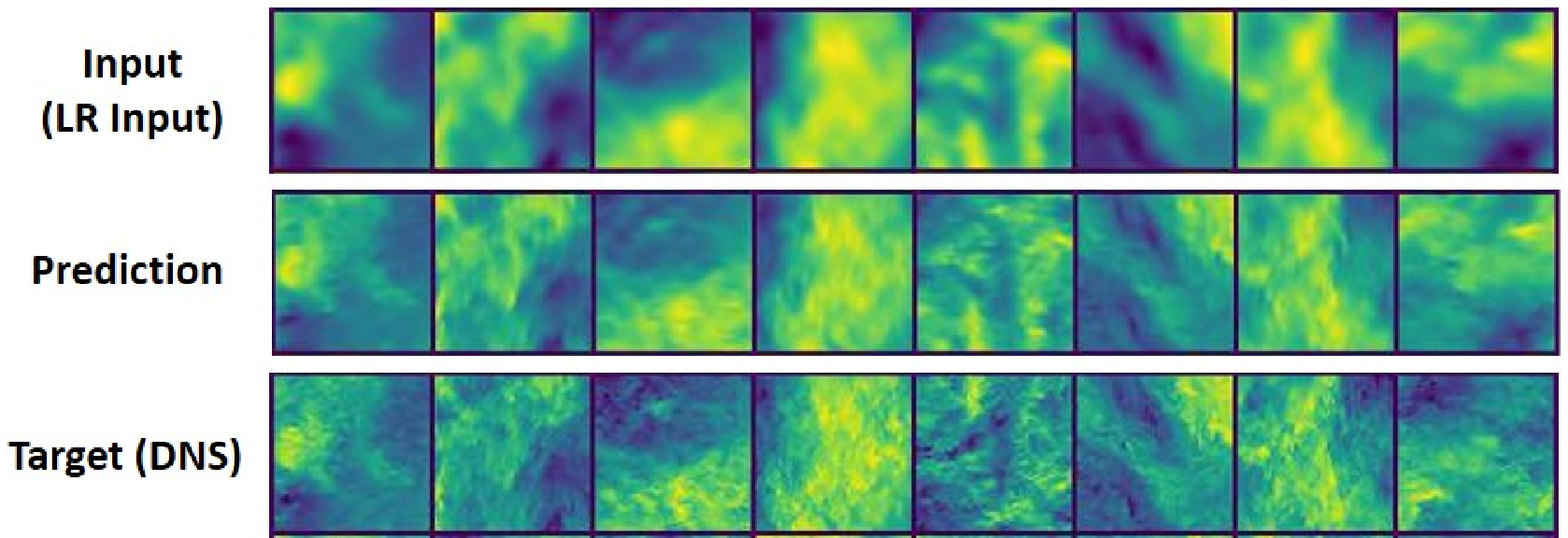}}
  \caption{Comparison of U-velocity reconstruction results on the 3 different Low-resolution Test Datasets. a)LR-Fine Dataset; b) LR-Medium Dataset; c) LR-Coarse Dataset}
\label{fig:result-pics}
\end{figure}
\begin{figure}
  \centerline{\includegraphics[width=5in]{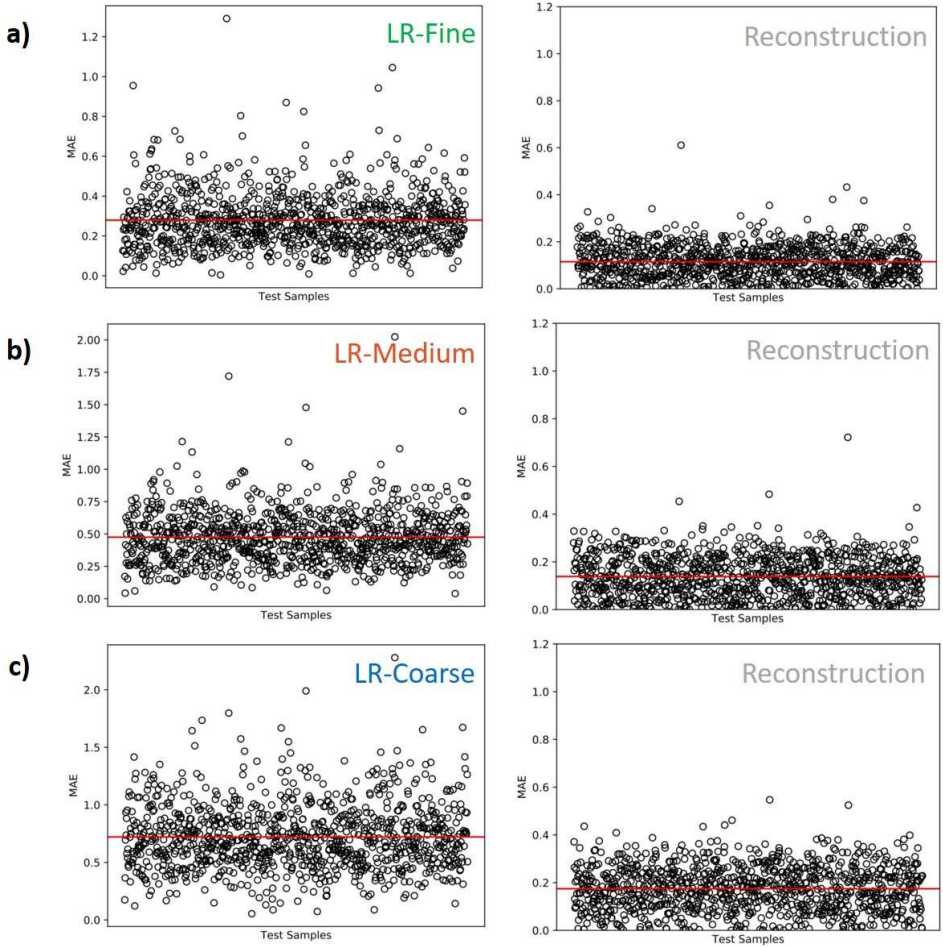}
  }
  \caption{MAE of Kinetic Energy for LR(left) and Reconstruction(right) Test Datasets. a)LR-Fine Dataset; b) LR-Medium Dataset; c) LR-Coarse Dataset. Notice the reduction in MAE on the reconstruction plots compared to low-resolution inputs}
  \label{fig:KE}
\end{figure}

In terms of fluid dynamics based metrics, there is good agreement between the DNS and the reconstructed values. On unseen test datasets, our model only slightly under predicts the kinetic energy and turbulent velocity values (\textit{figure} \ref{fig:KE}, \textit{figure} \ref{fig:turbvel}) while showing marked improvement in its correlation with the ground truth when compared to the LR inputs. 
Also, the vorticity distribution plots show good agreement between the reconstruction and the ground truth values on all three datasets. From \textit{figure} \ref{fig:vort}, it can be seen that the probability distribution functions of vorticity on the LR-Fine dataset almost exactly overlaps with the ground truth values. The reconstruction on LR-Medium is also decent while the probability distribution on the LR-coarse dataset doesn't capture the fringes of the distribution (\textit{figure} \ref{fig:vort}) which is to be expected. Additionally, the distribution of the average turbulent velocity (\textit{figure}\ref{fig:turbvel}) yielded similar trends with best reconstruction results on the fine LR dataset followed by medium LR and coarse LR. Furthermore, even on visual inspection of the up-sampling, differences between the reconstruction and the ground truth are almost imperceptible for the LR-Fine dataset (\textit{figure} \ref{fig:result-pics}). However, slight differences become more and more apparent as the low-resolution input decreases in its level of refinement. Nonetheless, SR-DNS Net is able to inject meaningful sub-grid scale refinement even on the very coarse LR input data.
Also, by using the SR-DNS Net framework, we observe a significant performance improvement in terms of training and inference times (\textit{table}\ref{tab:perf}) when compared to ResNet based CNNs and GAN based models.
Compared to the implementation of Fukami et al. \citep{fukami2018super}, we were able to achieve better mean squared error reconstruction results while using only $1/4^{th}$ the inference time and a $1/3^{rd}$ of the training time of their model. Moreover, this reduction in computational run-times was achieved while using 4 times more data to train the model (10k v/s 40k). 
Finally, in order to significantly reduce the training times of our model we explored the possibility of implementing transfer learning. In neural networks, transfer learning refers to the prior initialization of network weights by adopting the learned weights of a previously solved similar problem. Previous works such as \citep{Guastoni_2020} have successfully utilized this technique to speed up the training times of models by factor of 4. Using an approach similar to aforementioned reference, we were able to train the model on both the LR-Coarse and LR-Fine datasets using the weights of a previously trained model on the LR-Medium dataset. By performing transfer learning we consistently saw 3-4x speed up (\textit{figure} \ref{loss_plot} and \textit{table} \ref{tab:perf}) in our training times which is in accordance to the findings of \citep{Guastoni_2020}.
\begin{table*}
  \begin{center}
\def~{\hphantom{0}}
 \begin{tabular}{p{0.125\linewidth}p{0.2\linewidth}p{0.15\linewidth}p{0.2\linewidth}p{0.25\linewidth}} \\
 \toprule
 \textbf{Mode} & \textbf{CNN (ResNet)}& \textbf{ESRGAN} & \textbf{SR-DNS Net} & \textbf{SR-DNS Net w/ Transfer Learning} \\ [0.5ex] 
 \hline
 Training & 9.71 hrs. & Pre-trained & 2.05 hrs. & 0.62 hrs. \\ 
  &  (368s/ep. x 95 ep.) &  & (185s/ep. x 40 ep.) & (185s/ep. x 12 ep.)\\ 
 \hline
 Inference & 5.39 x $10^{-3}$s & 3.76 x $10^{-2}$s & 3.07 x $10^{-3}$s & 3.07 x $10^{-3}$s \\  
 \bottomrule
\end{tabular}
  \caption{Comparison of training and inference times between a CNN ResNet, ESRGAN, SR-DNS Net and SR-DNS Net with Transfer Learning. Training is performed on 50k images of the LR-Medium dataset. Inference times are evaluated on a single image. A pre-trained ESRGAN model with 4x upscaling was used to compare reconstruction performance during inference}
  \label{tab:perf}
  \end{center}
\end{table*}
\begin{figure}
  \centerline{
  \includegraphics[width=3in]{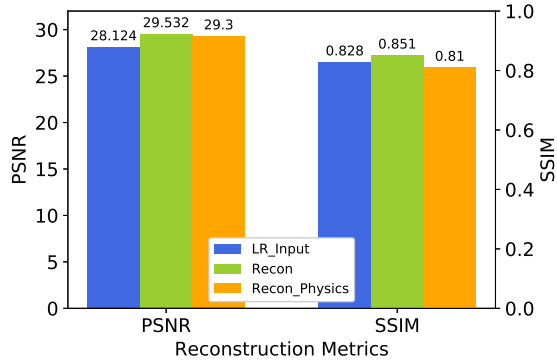}}
  \caption{Comparison of reconstruction metrics for the Reconstruction Loss v/s Reconstruction + Physics Loss ($\gamma = 0.01$). The physics based loss gives inferior results compared to L1 loss}
  \label{fig:phys_bar}
\end{figure}
\begin{figure}
     \centering
     \subfloat[][LR-Fine Dataset]{\includegraphics[width=0.45\linewidth]{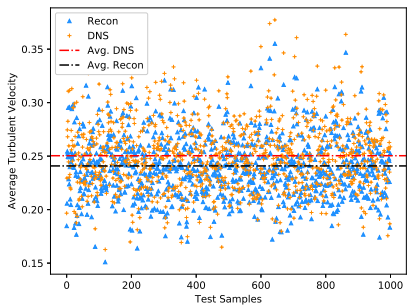}} \hfill
     \subfloat[][LR-Medium Dataset]{\includegraphics[width=0.45\linewidth]{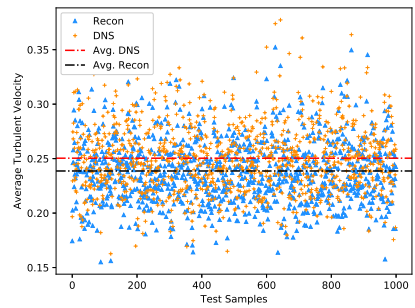}}\hfill
     \subfloat[][LR-Coarse Dataset]{\includegraphics[width=0.45\linewidth]{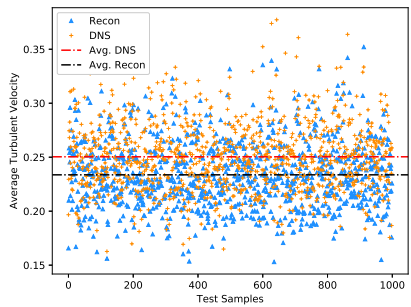}}
      \caption{Comparison of average turbulent velocity Distribution between reconstruction and the ground truth(DNS) on the test datasets. Notice the high degree of overlap beween the blue and orange markers along with slight offset in the average values}
      \label{fig:turbvel}
\end{figure}
\begin{figure}
  \centerline{
  \includegraphics[width=4.5in]{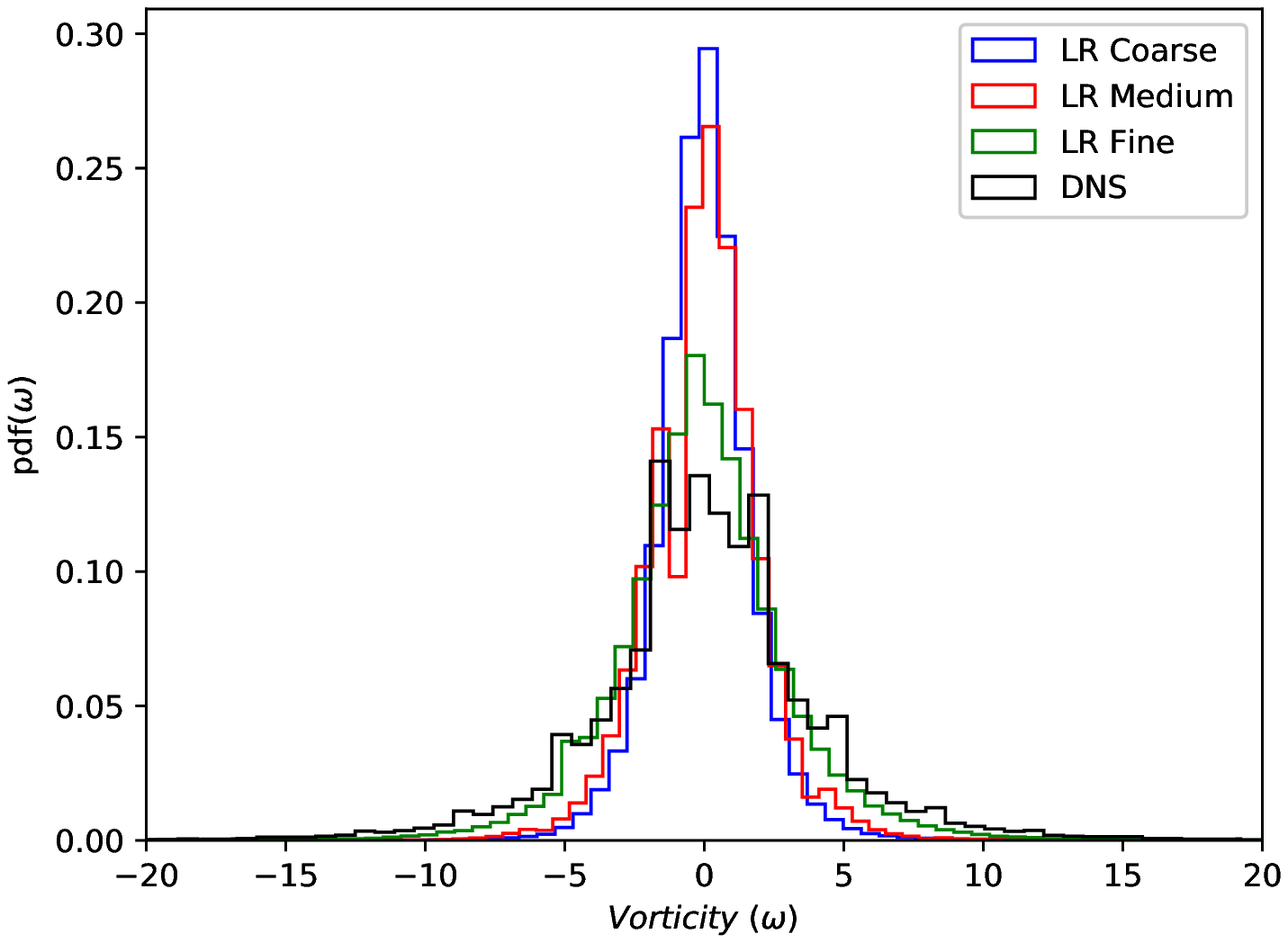}}
  \caption{Comparison of vorticity probability distribution function of the ground truth(DNS) with reconstruction on the different test sets}
  \label{fig:vort}
\end{figure}

To further enhance the interpretability and the fidelity of our framework, we experimented with backpropagating on physics based losses. Thus, we tried out a novel loss formulation consisting of two terms: 1. Reconstruction Loss (L1 Loss) and 2. Physics Loss . The physics loss in itself comprised of two components. Kinetic energy(K.E. Loss) and the continuity(Continuity Loss) (Equation \ref{eqn:loss}).
Therefore, the physics loss provided input to the model based on the conservation of kinetic energy and mass. In order to ensure balanced training of the network in terms of simultaneously minimizing both reconstruction and physics losses we introduced a hyperparameter term $\gamma$ to provide appropriate scaling. By incorporating a physics-based loss we saw a slight reduction in overall reconstruction performance \textit{figure} \ref{fig:phys_bar}. Also, the model convergence times were adversely affected. Additionally, by performing a grid search, we observed that the model performance worsened as the contribution of the physics loss ($\gamma$) was increased, with the best performance achieved when $\gamma$ = 0. We believe that this can be attributed to the strong regularization imposed by the additional loss terms. Another, important consideration is that by minimizing the velocity reconstruction loss we are already implicitly reducing the K.E. Loss and Continuity Losses. Thus, adding the physics based loss term doesn't add any new information into the model while simultaneously adding a regularization term that increases the model complexity, thereby resulting in inferior performance. 

\begin{equation}
    \mathbf{L}= \mathbf{L}_{Recon.} + \gamma (\mathbf{L}_{K.E.} + \mathbf{L}_{Continuity})
    \label{eqn:loss}
\end{equation}

\section{Conclusions}
A deep learning model (SR-DNS Net) has been developed to perform super-resolution based reconstruction of turbulent DNS data from low-resolution LES like data. 
To achieve this we have created a novel architecture (SR-DNS Net) which combines state-of-the-art ML building blocks such as MobileNets, U-Nets and Pixel-Shuffle layers to yield a highly computationally efficient super-resolution framework.
Our model yields a significant improvement in terms of image similarity metrics (PSNR and SSIM) between snapshots of the reconstruction and the low-resolution input. Furthermore, the model provides a highly accurate reconstruction of key flow metrics such as turbulent velocity, vorticity and kinetic energy for the isotropic homogeneous turbulence test case. To further validate the model's efficacy, it has been tested on 3 datasets with varying levels of coarseness (LR-Fine, LR-Medium and LR-Coarse). 
Using SR-DNS Net, a mapping between the coarser LR data and the refined DNS data was successfully learned. The application of this learned mapping to LR data yielded a marked improvement to the result terms of similarity with the ground truth. 
This learning can thus be subsequently injected into unseen low-resolution CFD snapshots (test data) to reconstruct the unfiltered, high-frequency sub-grid scale features. 
Therefore, using the DNS reconstructions, we can extract statistical insights and sub-grid scale features that would otherwise be absent from the low-resolution solutions. Our framework thereby facilitates the use of coarser grids by upscaling low-resolution, low-fidelity CFD snapshots. 
Additionally, our network forms a highly computationally efficient deep learning architecture which greatly improves upon training and inference times. Compared to previous implementations, \cite{fukami2018super}, we were able to achieve better mean squared error reconstruction results while using only a fraction of the inference time and training times. Moreover, this reduction in computational run-times was achieved while using 4 times more data to train the model (10k v/s 40k). 
We further enhance the model's computational efficiency by performing transfer learning which greatly accelerates model convergence during training by a factor of around 3-4x. 
Through this work, we also investigated the effect of adding physics based losses (kinetic energy and continuity losses) to the overall model performance. By analyzing the difference in results, we concluded that adding physics based losses yielded inferior results when compared to the conventional norm based reconstruction metrics. 

Since our current framework doesn't satisfy of the governing Navier-Stokes Equations or guarantee numerical stability the upsampled data might not be suitable for direct use in CFD solvers. However, models such as SR-DNS Net can still yield helpful statistical insights into high resolution fluid flow metrics such as vorticity, turbulent velocity and kinetic energy.
Furthermore, we hope that our research serves as an important initial step towards demonstrating the effectiveness and future potential of such super-resolution based reconstruction models. 
By incorporating models that build upon SR-DNS Net, super-resolution augmented CFD solvers will have the potential to greatly reduce the massive computational overhead usually associated with high fidelity CFD solutions. 
Such models would be able to directly map the LES like LR solution directly to HR DNS and can potentially eliminate the need for sub-optimal empirical models for sub-grid scale turbulence.


Finally, A physics informed super-resolution model that builds upon SR-DNS Net could be used in the future to perform software-in-the-loop upscaling inside CFD software. 
Such models would efficiently increase solution fidelity of CFD simulations while mostly mitigating the associated computational costs. 

\section{Acknowledgements}

This work is supported by the start-up fund provided by CMU Mechanical Engineering.

\section{Declaration of Interests}
The authors report no conflict of interest.

\bibliographystyle{jfm}
\bibliography{jfm-instructions}

\end{document}